\documentclass[aps,prl,noeprint,twocolumn,showpacs,amsmath,amssymb]{revtex4-2}
\usepackage{amsmath}
\usepackage{graphicx}
\usepackage{subfigure}
\usepackage{epstopdf}
\usepackage{color}
\usepackage{multirow}
\usepackage{setspace}
\usepackage{overpic}
\usepackage{amssymb}
\usepackage[bookmarksnumbered, pdfstartview=FitH,colorlinks,urlcolor=blue, citecolor=blue,linkcolor=blue]{hyperref}
\usepackage{lineno}
\usepackage{bm}
\usepackage{rotating}
\usepackage[utf8]{inputenc}
\let\oldequation\equation
\let\oldendequation\endequation

\renewenvironment{equation}
  {\linenomathNonumbers\oldequation}
  {\oldendequation\endlinenomath}

\begin{document}

\title{\bf \boldmath
First Measurement of the Decay Asymmetry in the pure \textit{W}-boson-exchange Decay \texorpdfstring{$\Lambda_{c}^{+}\to\Xi^{0}K^{+}$}{Lc2Xi0K}}

\author{
\begin{small}
\begin{center}
M.~Ablikim,$^{1}$ M.~N.~Achasov,$^{5,b}$ P.~Adlarson,$^{74}$ X.~C.~Ai,$^{80}$ R.~Aliberti,$^{35}$ A.~Amoroso,$^{73a,73c}$ M.~R.~An,$^{39}$ Q.~An,$^{70,57}$ Y.~Bai,$^{56}$ O.~Bakina,$^{36}$
% ...
\\
\vspace{0.2cm}
(BESIII Collaboration)\\
\vspace{0.2cm} {\it
$^{1}$ Institute of High Energy Physics, Beijing 100049, People's Republic of China\\
% ...
$^{5}$ Budker Institute of Nuclear Physics SB RAS (BINP), Novosibirsk 630090, Russia\\
% ...
$^{35}$ Johannes Gutenberg University of Mainz, Johann-Joachim-Becher-Weg 45, D-55099 Mainz, Germany\\
$^{36}$ Joint Institute for Nuclear Research, 141980 Dubna, Moscow region, Russia\\
% ...
$^{39}$ Liaoning Normal University, Dalian 116029, People's Republic of China\\
% ...
$^{56}$ Southeast University, Nanjing 211100, People's Republic of China\\
$^{57}$ State Key Laboratory of Particle Detection and Electronics, Beijing 100049, Hefei 230026, People's Republic of China\\
% ...
$^{70}$ University of Science and Technology of China, Hefei 230026, People's Republic of China\\
% ...
$^{73a}$ University of Turin and INFN, University of Turin, I-10125, Turin, Italy\\
$^{73b}$ University of Eastern Piedmont, I-15121, Alessandria, Italy\\
$^{73c}$ INFN, I-10125, Turin, Italy\\
$^{74}$ Uppsala University, Box 516, SE-75120 Uppsala, Sweden\\
% ...
$^{80}$ Zhengzhou University, Zhengzhou 450001, People's Republic of China\\
\vspace{0.2cm}
% ...
$^{b}$ Also at the Novosibirsk State University, Novosibirsk, 630090, Russia\\
% ...
}\end{center}
\vspace{0.4cm}
\end{small}
}
\vspace{4cm}

\date{\it \small \bf \today}

\begin{abstract}
Based on $4.4~\text{fb}^{-1}$ of $e^{+}e^{-}$ annihilation data collected at the center-of-mass energies between $4.60$ and $4.70~\text{GeV}$ with the BESIII detector at the BEPCII collider, the pure \textit{W}-boson-exchange decay $\Lambda_{c}^{+}\to\Xi^{0}K^{+}$ is studied with a full angular analysis. 
The corresponding decay asymmetry is measured for the first time to be $\alpha_{\Xi^{0}K^{+}}=0.01\pm0.16({\rm stat.})\pm0.03({\rm syst.})$. This result reflects the non-interference effect between the $S$- and $P$-wave amplitudes. The phase shift between $S$- and $P$-wave amplitudes has two solutions, which are $\delta_{p}-\delta_{s}=-1.55\pm0.25({\rm stat.})\pm0.05({\rm syst.})~\text{rad}$ or $1.59\pm0.25({\rm stat.})\pm0.05({\rm syst.})~\text{rad}$.
\end{abstract}


\maketitle

\oddsidemargin  -0.2cm
\evensidemargin -0.2cm

Investigations of charmed baryon decay dynamics are essential to explore the weak and strong interactions in the Standard Model~(SM) 
of particle physics.
The ground state of the singly-charmed baryons $\Lambda_{c}^{+}$ was discovered in 1979~\cite{Abrams:1979iu}.
Many studies have since been made of the properties of charmed baryons, such as the decay branching fractions~(BFs) and decay asymmetries.
But experimental results of the decay asymmetries, which are sensitive to the different amplitudes in the decay dynamics, were only a few.
Since 2014, there has been some progress on the weak hadronic decays of $\Lambda_{c}^{+}$, $\Xi_c^{+,0}$, and $\Omega_c^0$, both experimentally and theoretically~\cite{Workman:2022ynf,Cheng:2021qpd,Cheng:2021vca}.
This provides crucial information about the properties of all the singly-charmed baryons and the searches for doubly-charmed baryons ($\Xi_{cc}$ and $\Omega_{cc}$)~\cite{Yu:2017zst}.
Nonetheless, the understanding of the decay dynamics of charmed baryons is still limited, due to the lack of precision experimental measurements and the difficulties in the theoretical treatment of strong interaction effects. 

\begin{figure}[htbp]\centering
	\includegraphics[width=0.60\linewidth]{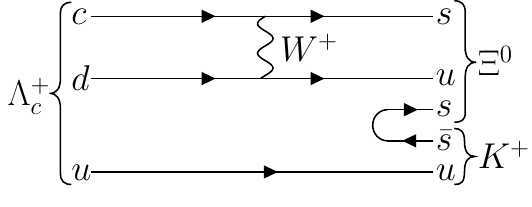}
	\includegraphics[width=0.60\linewidth]{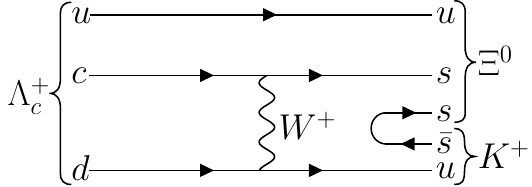}
	\caption{Feynman diagrams for $\Lambda_{c}^{+}\to\Xi^{0}K^{+}$.}
	\label{fig:feynman}
\end{figure}

Compared to heavy meson decays, charmed baryon decays have a significant dependence on nonfactorizable contributions from \textit{W}-boson-exchange diagrams. However, these contributions cannot currently be calculated using theoretical approaches. Additionally, no experimental measurements exist for the decay asymmetries of \textit{W}-boson-exchange hadronic decays. An example of such a process is the decay $\Lambda_{c}^{+}\to\Xi^{0}K^{+}$, which can only be produced via a \textit{W}-boson-exchange process as depicted in Fig.~\ref{fig:feynman}.
Experimental measurements of the asymmetry parameters of the decay $\Lambda_{c}^{+}\to\Xi^{0}K^{+}$ can aid understanding of the internal dynamics and can also explore charge-parity ($CP$) violation in baryons~\cite{Zou:2019kzq}.

Table~\ref{tab:sum} lists the theoretical calculations of the BF and asymmetry parameters 
of $\Lambda_{c}^{+}\to\Xi^{0}K^{+}$, as well as the experimental measurements of the BF. Various predictions of the BFs based on the covariant confined quark model (CCQM), the pole model, and current algebra (CA) in the 1990s are all smaller than the experimental results~\cite{Workman:2022ynf}.
This is explained as a strong cancellation in the $S$- and $P$-wave amplitudes, corresponding to the $L=0$ and $L=1 $ orbital angular momenta of the $\Xi^{0}$-$K^{+}$ system, respectively.
Moreover, the decay asymmetry parameter was predicted to be zero in these models owing to the vanishing $S$-wave amplitude~\cite{Korner:1992wi,Xu:1992vc,Zenczykowski:1993jm,Ivanov:1997ra,Sharma:1998rd}.
This long-standing puzzle has recently experienced a renewed interest in the theoretical
community~\cite{Zou:2019kzq,Geng:2019xbo,Zhong:2022exp}, especially after the report of new BF measured by BESIII in 2018~\cite{BESIII:2018cvs}. To reproduce the relatively large experimental branching fraction of this decay, Ref.~\cite{Zou:2019kzq} adopted a variant of the CA approach and obtained a larger $\mathcal{B}(\Lambda_c^+\to \Xi^0 K^+) \simeq 0.71\%$. This modification introduces a large positive decay asymmetry of $0.90$, which is quite close to the calculations based on SU(3) symmetry~\cite{Geng:2019xbo,Zhong:2022exp}.
However, regarding the significant enhancement of $\alpha_{\Xi^0 K^+}$ from 0 to about 0.9, the authors of Ref.~\cite{Groote:2021pxt} pointed out that the particular construction of the $S$-wave amplitude in Ref.~\cite{Zou:2019kzq} is not well-justified. So experimental measurement of the asymmetry parameter of the decay $\Lambda_{c}^{+}\to\Xi^{0}K^{+}$ will be crucial to test these calculations and confirm the vanishing $S$-wave contribution~\cite{Groote:2021pxt}.

\begin{table*}[htbp]
	\centering
	\caption{Theoretical calculations and experimental measurements of the BF, $\alpha_{\Xi^{0}K^{+}}$, $|A|$, $|B|$ and $\delta_{p}-\delta_{s}$ of $\Lambda_{c}^{+}\to\Xi^{0}K^{+}$. $G_{F}$ is the Fermi constant. The superscript $a$ denotes a model with SU(3) symmetry, while model $b$ includes SU(3) symmetry-breaking effects.  The PDG Fit BF also includes a CLEO result on $\mathcal{B}(\Lambda_{c}^+ \to \Xi^0K^+)/\mathcal{B}(\Lambda_{c}^+ \to p K^- \pi^+)$~\cite{Workman:2022ynf}.}
	\label{tab:sum}
	\begin{tabular}{c c c c c c}
		\hline\hline
  Theory or experiment & $\mathcal{B}(\Lambda_{c}^{+}\to\Xi^{0}K^{+})$  & \raisebox{0.2ex}{$\alpha_{\Xi^{0}K^{+}}$}
 & $|A|$    & $|B|$  & $\delta_{p}-\delta_{s}$\\
	               & ($\times10^{-3}$) &  & $(\times10^{-2}G_{F}~\text{GeV}^{2})$    & $(\times10^{-2}G_{F}~\text{GeV}^{2})$  & $(\text{rad})$\\
		\hline
		K{\"o}rner (1992), CCQM~\cite{Korner:1992wi}     & 2.6                  & 0                           & $\cdots$                   & $\cdots$                     & $\cdots$ \\
		
		Xu (1992), Pole~\cite{Xu:1992vc}                 & 1.0                  & 0                           & $0$                        & $7.94$                       & $\cdots$ \\
		
		{\'Z}encaykowski (1994), Pole~\cite{Zenczykowski:1993jm}& 3.6           & 0                           & $\cdots$                   & $\cdots$                     & $\cdots$ \\
		
		Ivanov (1998), CCQM~\cite{Ivanov:1997ra}         & 3.1                  & 0                           & $\cdots$                   & $\cdots$                     & $\cdots$ \\
		
		Sharma (1999), CA~\cite{Sharma:1998rd}           & 1.3                  & 0                           & $\cdots$                   & $\cdots$                     & $\cdots$ \\
		
		Geng (2019), SU(3)~\cite{Geng:2019xbo}           & $5.7\pm0.9$          & $0.94^{+0.06}_{-0.11}$      & $2.7\pm0.6$                & $16.1\pm2.6$                 & $\cdots$ \\
		
		Zou (2020), CA~\cite{Zou:2019kzq}                & $7.1$                & $0.90$                      & $4.48$                     & $12.10$                      & $\cdots$ \\

            Zhong (2022), SU(3)$^{a}$~\cite{Zhong:2022exp}   & $3.8^{+0.4}_{-0.5}$  & $0.91^{+0.03}_{-0.04}$      & $3.2\pm0.2$                & $8.7^{+0.6}_{-0.8}$          & $\cdots$ \\
		
		Zhong (2022), SU(3)$^{b}$~\cite{Zhong:2022exp}   & $5.0^{+0.6}_{-0.9}$  & $0.99\pm0.01$               & $3.3^{+0.5}_{-0.7}$        & $12.3^{+1.2}_{-1.8}$         & $\cdots$ \\
		\hline
		BESIII (2018)~\cite{BESIII:2018cvs}              & $5.90\pm0.86\pm0.39$ & $\cdots$                    & $\cdots$                   & $\cdots$                     & $\cdots$ \\
		
		PDG Fit (2022)~\cite{Workman:2022ynf}            & $5.5\pm0.7$          & $\cdots$                    & $\cdots$                   & $\cdots$                     & $\cdots$ \\
		\hline\hline
	\end{tabular}
\end{table*}

In the SM, the amplitude for a spin-$1/2$ baryon decaying into a spin-$1/2$ baryon and a spin-$0$ meson can be written as $\mathcal{M} = i \bar{u}_{f}(A-B\gamma_{5})u_{i}$, where $A$ and $B$ are constants, $u_{i}$ and $\bar{u}_{f}$ are spinors describing the initial and final baryons~\cite{Xu:1992vc}.  For the decay $\Lambda_{c}^{+}\to\Xi^{0}K^{+}$, the decay asymmetry is defined by $\alpha_{\Xi^{0}K^{+}}=2\text{Re}(s^{*}p)/(|s|^{2}+|p|^{2})$, where $s=A$ and $p=|\vec{p}_{\Xi^{0}}|B/(E_{\Xi^{0}}+m_{\Xi^{0}})$; here $E_{\Xi^{0}}$ and $\vec{p}_{\Xi^{0}}$ are the energy and momentum of the $\Xi^{0}$ in the $\Lambda_{c}^{+}$ rest frame~\cite{Workman:2022ynf}. 
The effect of the $S$- and $P$-wave phase shift difference, $\delta_{p}-\delta_{s}$, is not well accounted for in the theoretical calculations of decay asymmetries. It can be extracted from experiments combined with the BF cited from the Particle Data Group~(PDG)~\cite{Workman:2022ynf} and provides an important experimental parameter for the theoretical prediction of $CP$ violation~\cite{Wang:2022tcm}.

In this Letter, we present the first measurement of the decay asymmetry of $\Lambda_{c}^{+}\to\Xi^{0}K^{+}$ and its decay dynamic parameters ($|A|$, $|B|$, and $\delta_{p}-\delta_{s}$). A multidimensional angular analysis of the cascade-decay $e^{+}e^{-}\to\Lambda_{c}^{+}\bar{\Lambda}_{c}^{-}$, $\Lambda_{c}^{+}\to\Xi^{0}K^{+}$, $\Xi^{0}\to\Lambda\pi^0$, and $\Lambda\to p\pi^{-}$ is performed using a technique similar to that used to measure the asymmetry parameters of $\Lambda_{c}^{+}\to pK_{S}^{0}$, $\Lambda\pi^{+}$, $\Sigma^{+}\pi^{0}$, and $\Sigma^{0}\pi^{+}$~\cite{BESIII:2019odb}.
The data samples used in this analysis, with an integrated luminosity of $4.4~\text{fb}^{-1}$, were collected at center-of-mass (CM) energies of $4.60~(587~\text{pb}^{-1})$, $4.63~(522~\text{pb}^{-1})$, $4.64~(552~\text{pb}^{-1})$, $4.66~(529~\text{pb}^{-1})$, $4.68~(1667~\text{pb}^{-1})$, and $4.70~{\rm GeV}$~($536~\text{pb}^{-1}$) with the BESIII detector at the BEPCII collider~\cite{BESIII:2022dxl,BESIII:2022ulv,Ke:2023qzc}. The values in the parentheses are the corresponding luminosities. Details about BEPCII as well as BESIII and its sub-detectors can be found in Refs.~\cite{BESIII:2009fln,Yu:2016cof,BESIII:2020nme,Zhang:2022bdc,BESIII2017endcap}.
Large data samples taken around the $\Lambda^+_c\bar \Lambda_c^-$ production threshold allow to measure decay asymmetry of low BF decays with the single-tag technique~\cite{Li:2021iwf}. The low-background environment is more favorable to measure $\alpha_{\Xi^{0}K^{+}}$ accurately. 
Charge-conjugate modes are always implied unless explicitly stated otherwise.


Simulated event samples produced with a {\sc geant4}-based~\cite{GEANT4:2002zbu} Monte Carlo (MC) package, which provides the geometric description
of the BESIII detector and the detector response, are used to determine detection efficiencies and to estimate backgrounds. More details about simulations can be found in Ref.~\cite{BESIII:2022xne}. The phase space MC signal sample is generated uniformly over phase space, which is $e^{+}e^{-}\to\Lambda_c^{+}\bar{\Lambda}_c^{-}$ followed by $\Lambda_{c}^{+}\to\Xi^{0}K^{+}$, $\Xi^{0}\to\Lambda(\to p\pi^{-})\pi^0(\to\gamma\gamma)$ and $\bar{\Lambda}_{c}^{-}$ decaying inclusively.
For the signal MC sample, the signal process is generated by the helicity formalism using the decay asymmetry parameters measured in this work or cited from the PDG~\cite{Workman:2022ynf}. 

A detailed description of the selection criteria for charged tracks, showers, $\pi^{0}$, and $\Lambda$ candidates is provided in Ref.~\cite{BESIII:2022xne}. The only difference is the $\chi^{2}$ of vertex fit, which constrains the daughter tracks $p\pi^{-}$ from $\Lambda$ decays to a common originating vertex, is imposed to be less than 20 in order to better suppress the background.
The $\Xi^{0}$ candidates are formed by $\Lambda\pi^{0}$ combinations and the invariant mass $M_{\Lambda\pi^{0}}$ must be within the mass region $(1.30,1.33)~\text{GeV}/c^{2}$. The mass region is selected to be about three times the resolution.

\begin{figure}[htbp]\centering
	\includegraphics[width=0.9\linewidth]{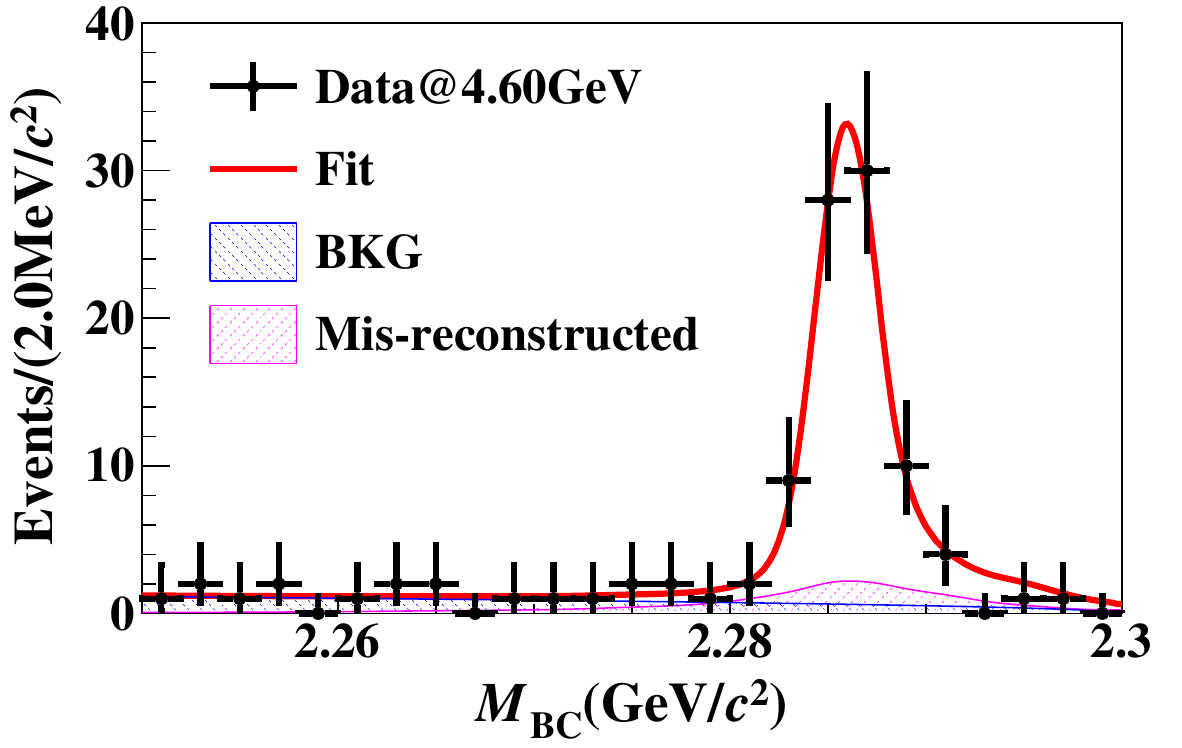}
	\caption{Distribution of the $M_{\text{BC}}$ fitting result at $4.60~{\rm GeV}$, and the corresponding signal yield is $70\pm8$. Black points with error bars are data; the blue shaded region indicates the combinatorial background events and pink shaded region is the mis-reconstructed signal events.}
	\label{fig:mBC_4600}
\end{figure}

Two kinematic variables, the energy difference $\Delta E\equiv E_{\Lambda_{c}^{+}}-E_{\text{beam}}$ and the beam-constrained mass $M_{\text{BC}}\equiv\sqrt{E_{\text{beam}}^{2}/c^{4}-|\vec{p}_{\Lambda_{c}^{+}}|^{2}/c^{2}}$, are defined to identify $\Lambda_{c}^{+}\to\Xi^{0}K^{+}$ candidates. 
Here, $E_{\Lambda_{c}^{+}}$ and $\vec{p}_{\Lambda_{c}^{+}}$ are the reconstructed energy and momentum of the $\Lambda_{c}^{+}$ candidates calculated in the $e^{+}e^{-}$ rest frame, and $E_{\text{beam}}$ is the average energy of the $e^{+}$ and $e^{-}$ beams. 
All candidates are required to satisfy $|\Delta E|<0.05~\text{GeV}$ and $2.25~\text{GeV}/c^{2}<M_{\text{BC}}< E_{\text{beam}}$. 
If more than one candidate satisfies the above requirements, the one with minimal $|\Delta E|$ is kept.
After applying these conditions, the $M_{\text{BC}}$ distribution in data collected at $4.60~{\rm GeV}$ is shown in the Fig.~\ref{fig:mBC_4600}. In the fit to this data, the correctly and mis-reconstructed signal shapes are modeled with the MC-simulated shape convolved with a Gaussian function representing the resolution difference between data and MC simulation, and the background shape is described by an ARGUS function~\cite{ARGUS:1990hfq}. 
Finally, $378\pm21$ signal events are obtained by combining the six energy points.

\begin{figure}[htbp]\centering
	\includegraphics[width=\linewidth]{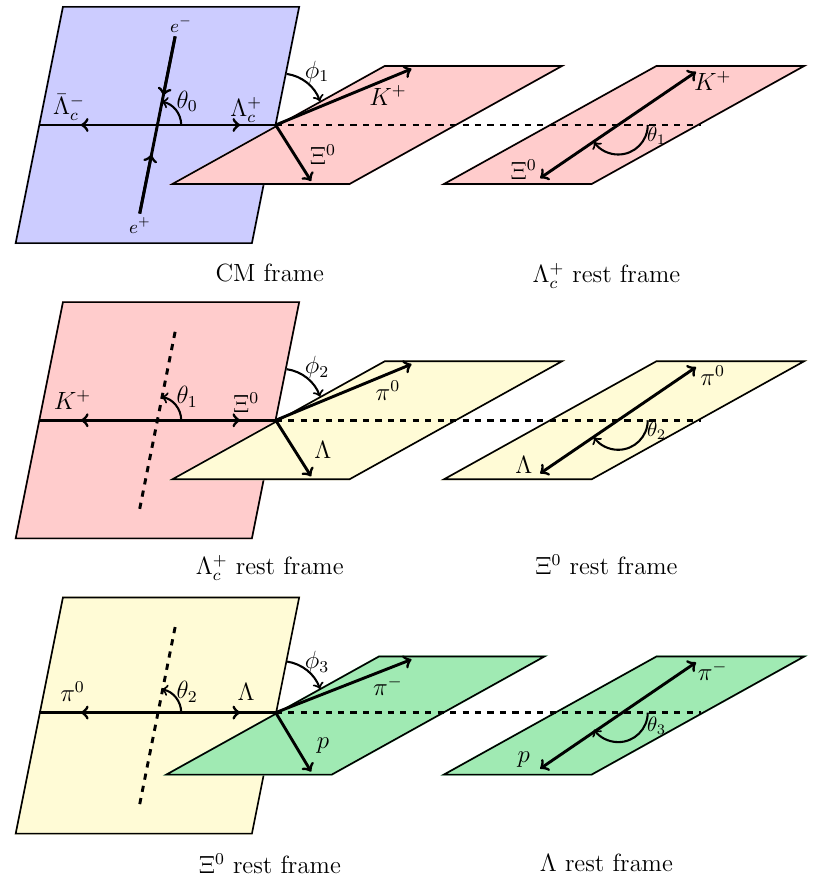}
	\caption{Definitions of the helicity frames and related angles for $e^{+}e^{-}\to\Lambda_{c}^{+}\bar{\Lambda}_c^{-}, \Lambda_{c}^{+}\to\Xi^{0}K^{+}, \Xi^{0}\to\Lambda\pi^0$, and $\Lambda\to p\pi^{-}$.}
	\label{fig:topo}
\end{figure}

The decay asymmetry parameters are determined by analyzing the multi-dimensional angular distributions, where the full cascade-decay chain is considered. 
The joint angular formula is obtained using the helicity basis~\cite{Supplemental}.
Figure~\ref{fig:topo} illustrates the definitions of the helicity angles for the three-level cascade decay $\Lambda_{c}^{+}\to\Xi^{0}K^{+}$, $\Xi^{0}\to\Lambda\pi^0$, and $\Lambda\to p\pi^{-}$ following the process of $e^{+}e^{-}\to\gamma^{*}\to\Lambda_{c}^{+}\bar{\Lambda}_c^{-}$.
In the helicity frame of the $e^{+}e^{-}\to\Lambda_{c}^{+}\bar{\Lambda}_c^{-}$, $\theta_{0}$ is the polar angle of the $\Lambda_{c}^{+}$ with respect to the $e^{+}$ beam axis in the $e^{+}e^{-}$ CM system. 
For the $\Lambda_{c}^{+}\to\Xi^{0}K^{+}$ decay, $\phi_{1}$ is the angle between the $e^{+}\Lambda_{c}^{+}$ and $\Xi^{0}K^{+}$ planes, and 
$\theta_{1}$ is the polar angle of the $\Xi^{0}$ with respect to the direction of $\bar{\Lambda}_{c}^{-}$ evaluated in $\Lambda_{c}^{+}$'s rest frame. For the $\Xi^{0}\to\Lambda\pi^0$ decay, $\phi_{2}$ is the angle between the $\Xi^{0}K^{+}$ and $\Lambda\pi^0$ planes, and $\theta_{2}$ is the polar angle of the $\Lambda$ with respect to the direction of $K^{+}$ evaluated in $\Xi^{0}$'s rest frame. For the helicity angles describing the $\Lambda\to p\pi^{-}$ decay, $\phi_{3}$ is the angle between the $\Lambda\pi^0$ and $p\pi^{-}$ planes, and $\theta_{3}$ is the polar angle of the proton with respect to the direction of $\pi^{0}$ evaluated in $\Lambda$'s rest frame.

In Ref.~\cite{Supplemental}, $\Delta_{0}$ is defined as the phase shift between two individual helicity amplitudes, $\mathcal{H}_{\lambda_{1},\lambda_{2}}$, for the $\Lambda_{c}^{+}$ production process $\gamma^{*}(\lambda_{0})\to\Lambda_{c}^{+}(\lambda_{1})\bar{\Lambda}_c^{-}(\lambda_{2})$ with $\gamma^{*}$'s helicity $\lambda_{0}=\pm1$, and total helicities $|\lambda_{1}-\lambda_{2}|=0$ and $1$, respectively. 
In the case where one-photon exchange dominates the production process, $\Delta_{0}$ is also the phase between the electric and magnetic form factors of $\Lambda_{c}^{+}$~\cite{Faldt:2017yqt,BESIII:2017kqg}, and $\alpha_{0}$ is the angular distribution parameter of $\Lambda_{c}^{+}$ defined by the helicity amplitude $\alpha_{0}=(|\mathcal{H}_{1/2,-1/2}|^{2} - 2\,|\mathcal{H}_{1/2,1/2}|^{2})/(|\mathcal{H}_{1/2,-1/2}|^{2} + 2\,|\mathcal{H}_{1/2,1/2}|^{2})$. 
Similarly, the $\Lambda_{c}^{+}\to\Xi^{0}K^{+}$ decay is described by two parameters, $\alpha_{\Xi^{0}K^{+}}$ and $\Delta_{\Xi^{0}K^{+}}$, where the latter one is the phase shift between the two helicity amplitudes. 
The Lee-Yang parameters~\cite{Supplemental,Lee:1957qs} can be obtained with the relations
\begin{equation}
	\label{relation}
	\begin{split}
		\beta_{\Xi^{0}K^{+}}&=\sqrt{1-(\alpha_{\Xi^{0}K^{+}})^{2}}~\text{sin}\Delta_{\Xi^{0}K^{+}},\\
		\gamma_{\Xi^{0}K^{+}}&=\sqrt{1-(\alpha_{\Xi^{0}K^{+}})^{2}}~\text{cos}\Delta_{\Xi^{0}K^{+}}.\\
	\end{split}
\end{equation}

In this analysis, the common free parameters ($\alpha_{\Xi^{0}K^{+}}$ and $\Delta_{\Xi^{0}K^{+}}$) describing the angular distributions for the six data sets are determined by a simultaneous unbinned maximum likelihood fit. 
The likelihood function is constructed from the joint probability density function (PDF) by
\begin{equation}
    \label{likelihood}
    \mathcal{L}_{\text{total}}=\sum^{\text{energy}}\mathcal{L}_{\text{data}}=\prod_{i=1}^{N_{\text{data}}}f_{s}(\vec\xi_{i}).
\end{equation}
Here, $f_{s}(\vec\xi_{i})$ is the PDF of the signal process, $N_{\text{data}}$ is the number of events in the data and $i$ is the event index. The signal PDF $f_{s}(\vec\xi_{i})$ is formulated as
\begin{equation}
	\label{signalPDF}
	f_{s}(\vec\xi_{i})=\frac{\epsilon(\vec\xi_{i})|M(\vec\xi_{i};\vec\eta)|^{2}}{\int\epsilon(\vec\xi_{i})|M(\vec\xi_{i};\vec\eta)|^{2}\text{d}\vec\xi_{i}}~,
\end{equation}
where $\vec\xi_{i}$ denotes the kinematic angular observables ($\theta_{0,1,2,3}$ and $\phi_{1,2,3}$) and $\vec\eta$ denotes the free parameters ($\alpha_{\Xi^{0}K^{+}}$ and $\Delta_{\Xi^{0}K^{+}}$) to be determined. 
$M(\vec\xi_{i};\vec\eta)$ is the total amplitude~\cite{Supplemental} of all decay chain and $\epsilon(\vec\xi_{i})$ is the detection efficiency parameterized in terms of the kinematic variables $\vec\xi_{i}$. 
The background contribution to the joint likelihood is subtracted according to the calculated likelihoods for the combinatorial background based on the inclusive MC simulation and for the mis-reconstructed signal events based on the signal MC simulation. 
The integration of the normalization factor is calculated with a large phase space MC sample as
\begin{equation}
	\label{integration}
\int\epsilon(\vec\xi_{i})|M(\vec\xi_{i};\vec\eta)|^{2}\text{d}\vec\xi_{i}=\frac{1}{N_{\text{gen}}}\sum_{k_{\text{MC}}}^{N_{\text{MC}}}|M(\vec\xi_{k_{\text{MC}}};\vec\eta)|^{2}~,
\end{equation}
where $N_{\text{gen}}$ is the total number of the simulated phase space MC events, $N_{\text{MC}}$ is the number of the phase space MC events surviving all selection criteria and $k_{\text{MC}}$ is the event index.

\begin{figure}[htbp]
	\centering
        \begin{overpic}[width=\linewidth]{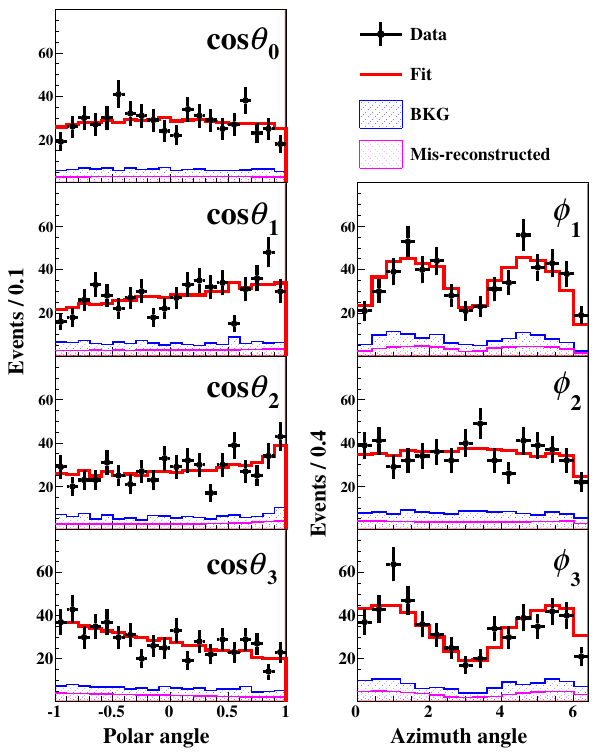}
        \end{overpic}
	\caption{Projections of the best fit onto various variables. Black points with error bars are data; red solid lines are phase space MC events re-weighted by angular distribution formula, and represent the fitting result; the blue shaded region denotes the combinatorial background events and the pink shaded region is the mis-reconstructed signal events.
}
	\label{anglefit}
\end{figure}

Using the MINUIT package~\cite{James:1975dr}, we minimize the negative logarithmic likelihood with background subtraction over the six data samples.
Here, $\alpha_{0}$, $\Delta_{0}$, $\alpha_{\Lambda\pi^{0}}$, $\alpha_{\bar{\Lambda}\pi^{0}}$, $\Delta_{\Lambda\pi^{0}}$, and 
 $\Delta_{\bar{\Lambda}\pi^{0}}$ are fixed to individual values measured by BESIII~\cite{BESIII:2023rwv,BESIII:2023drj}, where the $\alpha_{0}$ and $\Delta_{0}$ can be different at different energy points.  $\alpha_{p\pi^{-}}$ and $\alpha_{\bar{p}\pi^{+}}$ are fixed to the values from the PDG~\cite{Workman:2022ynf}.
In the fit, $\alpha_{\Xi^{0}K^{+}}$ and $\Delta_{\Xi^{0}K^{+}}$ are free parameters, and $\alpha_{\Xi^{0}K^{+}}=-\alpha_{\bar{\Xi}^{0}K^{-}}$ and $\Delta_{\Xi^{0}K^{+}}=-\Delta_{\bar{\Xi}^{0}K^{-}}$ as required under the $CP$ invariance assumption. 
The projections of the best fit onto several variables are shown in Fig.~\ref{anglefit}. 
The data are compared with the MC events weighted by the nominal fitting result.
From this fit, we obtain $\alpha_{\Xi^{0}K^{+}}=0.01\pm0.16$ and $\Delta_{\Xi^{0}K^{+}}=3.84\pm0.90~\text{rad}$.
Hence, the other two Lee-Yang parameters are calculated to be $\beta_{\Xi^{0}K^{+}}=-0.64\pm0.69$ and $\gamma_{\Xi^{0}K^{+}}=-0.77\pm0.58$, where the uncertainties are statistical only.

The systematic uncertainties arise mainly from the reconstruction of final state particles,  which is studied with $J/\psi\rightarrow K^{0}_{s}K^{\pm}\pi^{\mp}$ for kaon, $\Lambda^{+}_{c}\rightarrow\Lambda X$ for $\Lambda$, $\psi(3686)\rightarrow J/\psi \,\pi^{0}\pi^{0}$ and $e^{+}e^{-}\rightarrow\omega\pi^{0}$ for $\pi^{0}$. The systematic uncertainties for $\Delta E$ requirement and $M_{\text{BC}}$ signal regions are estimated by smearing the phase space MC samples using resolution parameters, and for the background subtraction by taking into account the background shape and size.The uncertainties from the quoted values of $\alpha_{0}$, $\Delta_{0}$, $\alpha_{\Lambda\pi^{0}}$, $\alpha_{\bar{\Lambda}\pi^{0}}$, $\Delta_{\Lambda\pi^{0}}$, $\Delta_{\bar{\Lambda}\pi^{0}}$, $\alpha_{p\pi^{-}}$ and $\alpha_{\bar{p}\pi^{+}}$ are estimated by Gaussian sampling considering their uncertainties and refit the angular distribution, and by taking the values in one time uncertainty of Gaussian fit as the uncertainties of this part. A further source of uncertainty is the fit bias, which is the difference before and after the correction from a pull distribution check.
Systematic uncertainties from all sources are combined in quadrature to calculate the total systematic uncertainties. All details of systematic uncertainties can be found in Ref.~\cite{Supplemental}.

\begin{figure}[htbp]\centering
	\includegraphics[width=\linewidth]{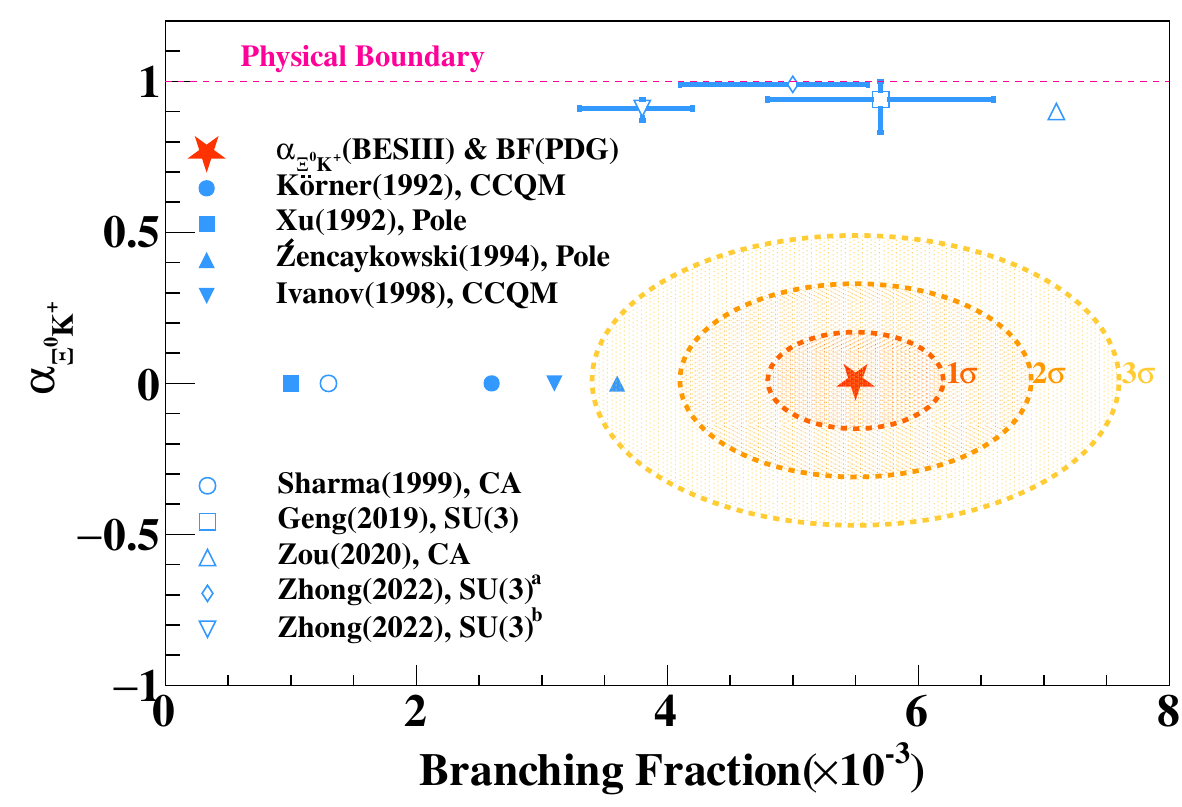}
	\caption{The comparison between this work and theoretical predictions, where the branching fraction is taken from PDG (2022)~\cite{Workman:2022ynf}. The $1\sigma$, $2\sigma$ and $3\sigma$ contours correspond to 68.2\%, 95.4\% and 99.7\% conference level, respectively. The blue symbols are theoretical predictions and the red star is result from this work. The definitions of the superscripts $a$ and $b$ can be found in Table~\ref{tab:sum} and the theory acronyms are explained in the text.}
	\label{res_compare}
\end{figure}

One has $\delta_{p}-\delta_{s}=\text{arctan}(\sqrt{1-\alpha^{2}_{\Xi^{0}K^{+}}}\text{sin}\Delta_{\Xi^{0}K^{+}}/\alpha_{\Xi^{0}K^{+}})$, and the derivation of $|A|$ and $|B|$ can be found in Ref.~\cite{Supplemental}. 
The study has uncovered two distinct physical solutions, with the first one characterized by $|A|=1.6^{+1.9}_{-1.6}({\rm stat.})\pm0.4({\rm syst.})$ and $|B|=18.3\pm2.8({\rm stat.})\pm0.7({\rm syst.})$, and the second one by $|A|=4.3\pm0.7({\rm stat.})\pm0.2({\rm syst.})$ and $|B|=6.7^{+8.3}_{-6.7}({\rm stat.})\pm1.6({\rm syst.})$.

In summary, by analyzing $4.4~\text{fb}^{-1}$ of $e^{+}e^{-}$ annihilation data collected at the CM energies between $4.60$ and $4.70~\text{GeV}$ with the BESIII detector, the pure \textit{W}-boson-exchange decay  $\Lambda_{c}^{+}\to\Xi^{0}K^{+}$ from the $e^{+}e^{-}\to\Lambda_{c}^{+}\bar{\Lambda}_{c}^{-}$ production has been studied.
The decay asymmetry parameters are measured for the first time as $\alpha_{\Xi^{0}K^{+}}=0.01\pm0.16({\rm stat.})\pm0.03({\rm syst.})$ and $\Delta_{\Xi^{0}K^{+}}=3.84\pm0.90({\rm stat.})\pm0.17({\rm syst.})~\text{rad}$.
The other two Lee-Yang parameters are calculated to be $\beta_{\Xi^{0}K^{+}}=-0.64\pm0.69({\rm stat.})\pm0.13({\rm syst.})$ and $\gamma_{\Xi^{0}K^{+}}=-0.77\pm0.58({\rm stat.})\pm0.11({\rm syst.})$.
The comparison between this work and theoretical predictions is shown in Fig.~\ref{res_compare}.
Our measurement of $\alpha_{\Xi^{0}K^{+}}$ is in good agreement with zero, which is consistent with the theoretical predictions from the 1990s.  
The decay dynamics parameters $|A|$, $|B|$, and $\delta_{p}-\delta_{s}$ are derived. The value of $\delta_{p}-\delta_{s}$ has two solutions, which are $\delta_{p}-\delta_{s}=-1.55\pm0.25({\rm stat.})\pm0.05({\rm syst.})~\text{rad}$ or $1.59\pm0.25({\rm stat.})\pm0.05({\rm syst.})~\text{rad}$.
This is of great significance for decay asymmetries, as $\text{cos}(\delta_{p}-\delta_{s})$ measured in this study is close to zero, an effect that had not been anticipated in previous literature.
This measurement resolves the long-standing puzzle and deepens our understanding of the strong dynamics in the charmed baryon sector.

P. R. Li and X. R. Lyu thank Hai-Yang Cheng, Fan-Rong Xu and Fu-Sheng Yu for useful discussions. The BESIII Collaboration thanks the staff of BEPCII and the IHEP computing center for their strong support. This work is supported in part by National Key R\&D Program of China under Contracts Nos. 2020YFA0406400, 2020YFA0406300; National Natural Science Foundation of China (NSFC) under Contracts Nos. 11635010, 11735014, 11835012, 11935015, 11935016, 11935018, 11961141012, 12022510, 12025502, 12035009, 12035013, 12061131003, 12192260, 12192261, 12192262, 12192263, 12192264, 12192265, 12221005, 12225509, 12235017; the Chinese Academy of Sciences (CAS) Large-Scale Scientific Facility Program; the CAS Center for Excellence in Particle Physics (CCEPP); Joint Large-Scale Scientific Facility Funds of the NSFC and CAS under Contract No. U1832207; CAS Key Research Program of Frontier Sciences under Contracts Nos. QYZDJ-SSW-SLH003, QYZDJ-SSW-SLH040; 100 Talents Program of CAS; Fundamental Research Funds for the Central Universities, Lanzhou University, University of Chinese Academy of Sciences; The Institute of Nuclear and Particle Physics (INPAC) and Shanghai Key Laboratory for Particle Physics and Cosmology; ERC under Contract No. 758462; European Union's Horizon 2020 research and innovation programme under Marie Sklodowska-Curie grant agreement under Contract No. 894790; German Research Foundation DFG under Contracts Nos. 443159800, 455635585, Collaborative Research Center CRC 1044, FOR5327, GRK 2149; Istituto Nazionale di Fisica Nucleare, Italy; Ministry of Development of Turkey under Contract No. DPT2006K-120470; National Research Foundation of Korea under Contract No. NRF-2022R1A2C1092335; National Science and Technology fund of Mongolia; National Science Research and Innovation Fund (NSRF) via the Program Management Unit for Human Resources \& Institutional Development, Research and Innovation of Thailand under Contract No. B16F640076; Polish National Science Centre under Contract No. 2019/35/O/ST2/02907; The Swedish Research Council; U. S. Department of Energy under Contract No. DE-FG02-05ER41374.


%

\end{document}